\documentclass[twocolumn,superscriptaddress,aps,prx]{revtex4-1}

\newcommand{\Tr}{\operatorname{Tr}}
\newcommand{\iu}{{i\mkern1mu}}
\newcommand*\diff{\mathop{}\!\mathrm{d}}

\usepackage{amssymb}
\usepackage{graphicx}
\usepackage{amsmath}
\usepackage{natbib}
\usepackage{siunitx}
\usepackage{textcomp}
\usepackage{bm}

\usepackage{color}

\usepackage{booktabs}
\usepackage{hyperref}

\begin{document}
	
	\title{Disconnecting a Traversable Wormhole: \\ Universal Quench Dynamics in Random Spin Models }
	\author{Tian-Gang Zhou}
	\affiliation{Institute for Advanced Study, Tsinghua University, Beijing,100084, China}
	\author{Lei Pan}
	\affiliation{Institute for Advanced Study, Tsinghua University, Beijing,100084, China}
	\author{Yu Chen}
	\affiliation{Graduate School of China Academy of Engineering Physics, Beijing, 100193, China}
	\author{Pengfei Zhang}
	\email{pengfeizhang.physics@gmail.com}
	\affiliation{Institute for Quantum Information and Matter, California Institute of Technology, Pasadena, California 91125, USA}
	\affiliation{Walter Burke Institute for Theoretical Physics, California Institute of Technology, Pasadena, California 91125, USA}
	\author{Hui Zhai}
	\email{hzhai@tsinghua.edu.cn}
	\affiliation{Institute for Advanced Study, Tsinghua University, Beijing,100084, China}
	\date{\today}
	
	\begin{abstract}
		
		Understanding strongly interacting quantum matter and quantum gravity are both important open issues in theoretical physics, and the holographic duality between quantum field theory and gravity theory nicely brings these two topics together. Nevertheless, direct connections between gravity physics and experimental observations in quantum matter are still rare. Here we utilize the gravity physics picture to understand quench dynamics experimentally observed in a class of random spin models realized in several different quantum systems, where the dynamics of magnetization are measured after the external polarization field is suddenly turned off. Two universal features of the magnetization dynamics, namely, a slow decay described by a stretched exponential function and an oscillatory behavior, are respectively found in different parameter regimes across different systems. This work addresses the issues of generic conditions under which these two universal features can occur, and we find that a natural answer to this question emerges in the gravity picture. By the holographic duality bridged by a model proposed by Maldacena and Qi, the quench dynamics after suddenly turning off the external polarization field is mapped to disconnecting an eternal traversable wormhole. Our studies show that insight from gravity physics can help unifying different experiments in quantum systems.         
		
	\end{abstract}
	
	\maketitle
	
	The holographic duality between quantum field theory at boundary and gravity theory in bulk has shed new insights in understanding both quantum matter and gravity \cite{holography1,holography2}. For example, on the gravity side, the wormhole in local Einstein gravity is not traversable, and recently mechanisms that can render a wormhole traversable have been studied by coupling quantum fields on boundaries \cite{Gao,traversable_WH,MQ}. On the quantum matter side, insights from gravity theory can help us understand strongly interacting quantum systems, such as non-Fermi liquids \cite{holography2}. Nevertheless, so far only a few specifically designed quantum models, such as the Sachdev-Ye-Kitaev (SYK) model \cite{SY,K1,K2,K3}, have been explicitly shown to possess holographic duality to gravity theory. Such a model is seemingly different from realistic systems in quantum materials and is also hard to be realized by quantum simulations. Therefore, so far little connection between sights from gravity physics and realistic experimental observations in quantum matter has been established through holographic duality, except for few known examples such as bounds for viscosity and transport coefficients \cite{bound1,bound2,bound3,bound4,bound5}.

	\begin{figure*}[tb]
		\centering
		\includegraphics[width=1.0\linewidth]{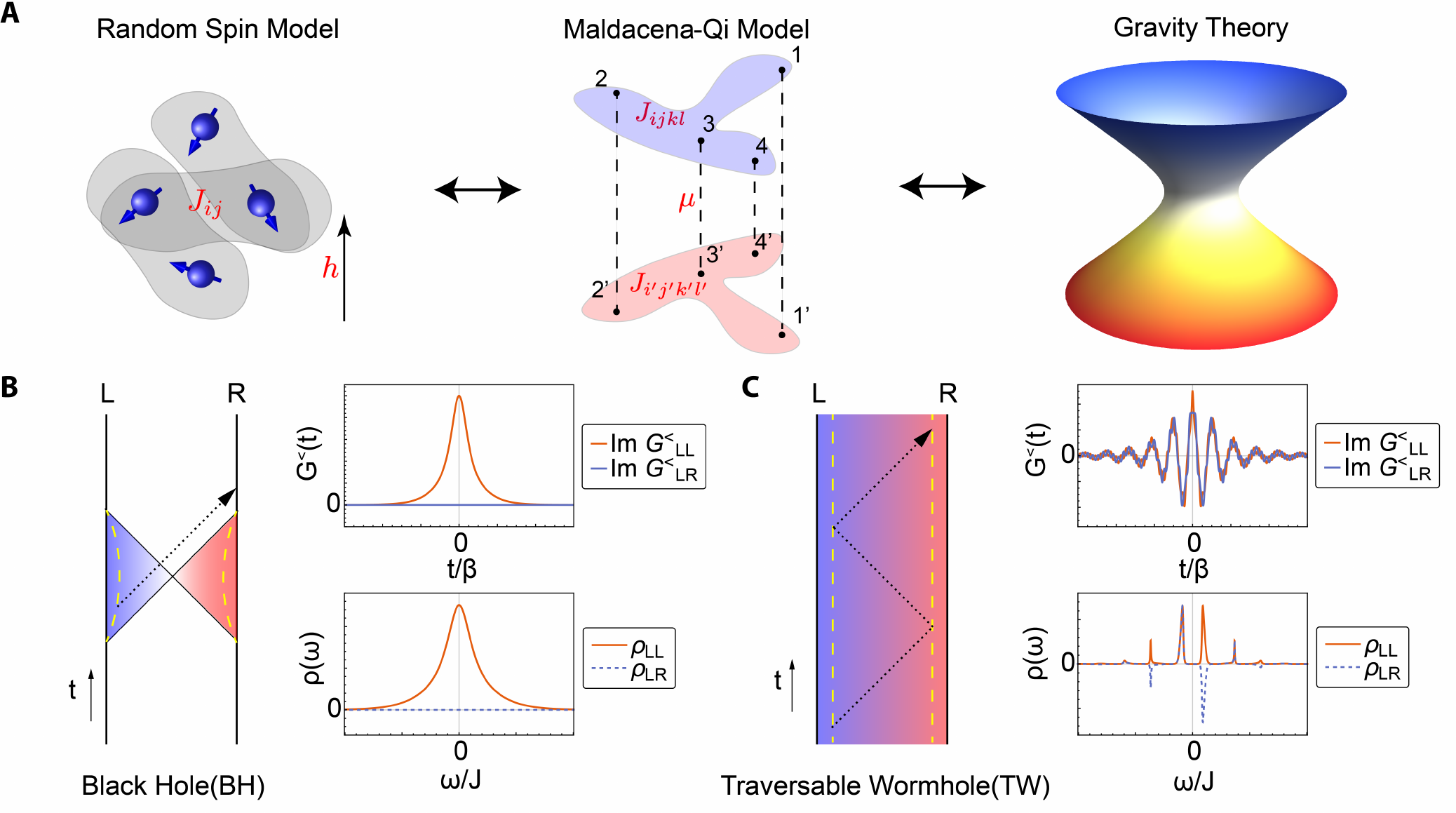}
		\caption{\textbf{Illustration of Models and Phases:} (A) The connection between the random spin models and the gravity theory bridged by the Maldacena-Qi model. (B-C) Illustration of the black hole (BH) phase and the eternal traversable wormhole (TW) phase in the Maldacena-Qi model and its dual gravity theory. The correspondence between the Penrose diagram in the gravity theory (left) and the lesser Green's function $G^{<}(t)$, as well as the spectral function $\rho(\omega)$ (right) in the quantum theory, are shown for the BH phase (B) and the TW phase (C), respectively. In the Penrose diagrams, the colored regimes are the physical space in the geometry. Two yellow dashed lines represent the physical boundaries of two $\text{AdS}_2$ spaces, denoted by the left (L) and the right (R) ones, where the quantum fields are defined. The black dotted lines with an arrow represent the null geodesic. The horizontal axis is the real-time axis.    } 
		\label{illustration}
	\end{figure*}

In this work we focus on a typical random spin model written as 
	\begin{equation}
	\hat{H}=\sum\limits_{i<j}J_{ij}(\hat{S}^x_i\hat{S}^x_j+\hat{S}^y_i\hat{S}^y_j+\Delta\hat{S}^z_i\hat{S}^z_j)-h \sum\limits_i \hat{S}^x_i. \label{random_spin}
	\end{equation} 
This model has recently been realized by cold molecules in optical lattices \cite{JILA13,JILA14}, and NV centers \cite{Harvard18}, fermions in harmonic traps \cite{Toronto19}, Rydberg atoms \cite{Heidelberg19}, high spin atoms in optical lattices \cite{Pairs20} and by solid state NMR \cite{NMR1,NMR2}. $J_{ij}$ is random all-to-all spin interaction coefficients, which originates either from dipolar interaction or from the spin exchanging interaction, and the randomness comes from the random locations of spin carriers in these systems. We can denote $J_{ij}=\bar{J}/N+\delta J_{ij}$, where $\bar{J}/N$ is the averaged value of all $J_{ij}$ and $N$ is the number of spins. $\langle \delta J^2_{ij}\rangle=4J^2/N$ and $J$ denotes the typical strength of random spin interactions. $\Delta$ is the anisotropy between the longitudinal and the transverse spin-spin interactions. $h$ is an external polarization field. Among these systems, $\hbar/J$ varies from a few microseconds to a few milliseconds, and the values of $\bar{J}/J$ and $\Delta$ are also different.  
	
Taking the advantages of the controllability of these systems, these realizations enable experimental studies of far-from-equilibrium quantum dynamics, and the quench dynamics in these random spin models have also been reported in a number of recent experiments \cite{JILA13,JILA14,Harvard18,Toronto19,Heidelberg19,Pairs20}. In the quench dynamics, spins are initially polarized in the transverse direction, say, along $\hat{x}$, by $h$-term in Eq. \ref{random_spin}, and then the dynamics of the total magnetization along $\hat{x}$ are experimentally measured after $h$ is suddenly turned off. In some of these experiments where the magnetization decay time due to the single-spin effect is shorter, a spin echo is applied to eliminate the single-spin effect \cite{JILA13,JILA14,Harvard18,Pairs20}, which ensures that the observed features are all due to spin-spin interactions described by Eq. \eqref{random_spin}. Two features of the quench spin dynamics have been observed in the time scale from a few to a few tens of $\hbar/J$. They take place in different parameter regimes, but both are observed in different systems and across different parameter regimes. These two features are listed below. 
	
	\begin{itemize}
		\item Slow decay of the total magnetization has been found when the magnetization decays to zero at long time, and the time dependence of the total magnetization is identified as a stretched exponential function.  This feature has been observed in experiments on NV centers \cite{Harvard18}, Rydberg atom \cite{Heidelberg19} and high spin atoms in optical lattices \cite{Pairs20}, 
		\item When the total magnetization saturates at finite value at long time, the total magnetization oscillates in time. This feature has been observed in experiments using cold molecules in optical lattices \cite{JILA13,JILA14}, and fermions in harmonic traps \cite{Toronto19}.
	\end{itemize}  
	 Here we will address the question of \textit{the general conditions under which these two features will occur in these random spin models.} This work is to show that a scenario emerged from the dual gravity theory unifies these experiments. 
	 
This work establishes a connection between the quench dynamics in these random spin models and the dynamics after suddenly turning off the coupling in a traversable wormhole. The connection is bridged by a model coupling two SYK systems proposed by Maldacena and Qi, as shown in Fig. \ref{illustration}(a). The Maldacena-Qi model is dual to a gravity theory hosting an eternal traversable wormhole \cite{MQ}, and a clear correspondence can be established between the behaviors of the spectral function in different phases of the Maldacena-Qi model and different space-time geometry in the gravity side with or without a traversable wormhole. We will show that the random spin models are closely related to the Maldacena-Qi model, and the spectral functions of these two models share similar behaviors. Hence, by studying the evolution of the spectral functions during the quench dynamics, we can establish a gravity interpretation of the quench dynamics.

	\vspace{0.05in}
	
	\textbf{Maldacena-Qi Model and its Gravity Dual.} Since the Maldacena-Qi model is a coupled SYK model, we need to first briefly introduce the SYK model. A single SYK model is written as
	\begin{equation}
	\hat{H}_{\text{SYK}}(\{\psi_i\})=\iu^2 \sum\limits_{1\leqslant i<j<k<l\leqslant N}J_{ijkl}\hat{\psi}_{i}\hat{\psi}_j\hat{\psi}_k\hat{\psi}_l, 
	\end{equation}
	where $\hat{\psi}_i (i=1,\dots,N)$ are $N$ Majorana fermion operators, and $J_{ijkl}$ are random Gaussian variables with zero mean and the variance $\langle J^2_{ijkl}\rangle=3! J^2/N^3$. The SYK model is exactly solvable in the large-$N$ limit. It can be shown explicitly that the low-energy physics of the SYK model has an emergent conformal symmetry, which can be dual to the Jackiw-Teitelboim gravity in the $\text{AdS}_2$ geometry with a black hole.  
	
	Various extensions of the SYK models have been considered, especially coupled SYK models \cite{coupledSYK1,coupledSYK2,coupledSYK3,coupledSYK4,coupledSYK5,coupledSYK6,coupledSYK7,coupledSYK8,coupledSYK9,coupledSYK10SymBreak,coupledSYK11SymBreak,coupledSYK12,MQ}. Here we focus on the version proposed by Maldacena and Qi \cite{MQ}. The Maldacena-Qi model is written as 
	\begin{equation}
	\hat{H}_{\text{MQ}}=\hat{H}_{\text{SYK}}(\{\hat\psi^L_i\})+\hat{H}_{\text{SYK}}(\{\hat\psi^R_i\})+i\mu\sum_{i}\hat\psi^\text{L}_i\hat\psi^{\text{R}}_i.
	\end{equation}
	where $\hat\psi^\text{L}_i$ and $\hat\psi^\text{R}_i$ denote the left and the right Majorana fermions. If $\mu=0$, there is no coupling between two SYK models, and on the gravity side, two $\text{AdS}_2$ spaces are also independent. Therefore, the geodesic starting from the boundary of one $\text{AdS}_2$ space can never reach the boundary of the other $\text{AdS}_2$ space. The $\mu$-term adds coupling between two boundaries of the $\text{AdS}_2$ spaces which distorts the boundary geometries. At finite temperature, when the distortion is weak, two boundaries still cannot be connected by a geodesic and this is referred to as the ``black hole" (BH) phase. However, when the distortion is strong enough, the geodesic starting from the boundary of one $\text{AdS}_2$ space can reach the boundary of the other $\text{AdS}_2$ space, which renders these two black holes into a traversable wormhole. This is referred to as the ``traversable wormhole" (TW) phase. The difference between these two geometries is represented by two different Penrose diagrams in Fig. \ref{illustration}(B) and (C). 
	
	In the presence of the holographic duality, the behavior of correlation functions are closely related to the connectivity of the space-time geometry through the standard holographic dictionary. Let us consider the lesser Green's function $G^{<}_{aa^\prime}(t)=i \sum_{l}\langle \hat\psi_l^{a^\prime}(t) \hat\psi_l^a(0) \rangle/N$, where $a,a^\prime=\text{L},\text{R}$. Roughly speaking, this Green's function can be physically interpreted as follows: first, a signal is created at $t=0$ by an operator at boundary $a^\prime$, and then the signal propagates following the geodesic, and finally, the signal is detected at time $t$ on boundary $a$. In the BH geometry, since there is no connection between two boundaries, it is natural to have $G^{<}_{aa^\prime}(t)=0$ for $a\neq a^\prime$. For $a=a^\prime$, $G^{<}_{aa^\prime}(t)$ decreases as $t$ increases, because as one can see from the Penrose diagram in Fig. \ref{illustration}(B), the geodesic, denoted by the black dotted line, moves away from the boundary. However, the situation is different for the TW geometry. In the TW geometry, as also shown by the Penrose diagram in Fig. \ref{illustration}(C), the geodesic starting from the left boundary can reach the right boundary and can be further bounced back and forth between two boundaries for an eternal traversable wormhole \cite{MQ}. Thus all $G^{<}_{aa^\prime}(t)$ are always finite for $a=a^\prime$ and $a\neq a^\prime$. At a time when the geodesic arrives at one of the boundaries, the corresponding Green's function reaches a maximum. Therefore, $G^{<}_{aa^\prime}(t)$ shows an oscillatory behavior in time. These two different behaviors of $G^{<}_{a a^\prime}(t)$ are also illustrated in Fig. \ref{illustration}(B) and (C). Similar behaviors can also be found in the quantity $i \sum_{l}\langle [\hat\psi_l^a(0), \hat\psi_l^{a^\prime}(t)] \rangle/N = - G^{>}_{aa^\prime}(t) - G^{<}_{aa^\prime}(t)$, which more directly reflects whether and how the information gets transferred from one side to the other.
	
	\begin{figure*}[tb]
		\centering
		\includegraphics[width=1.0\linewidth]{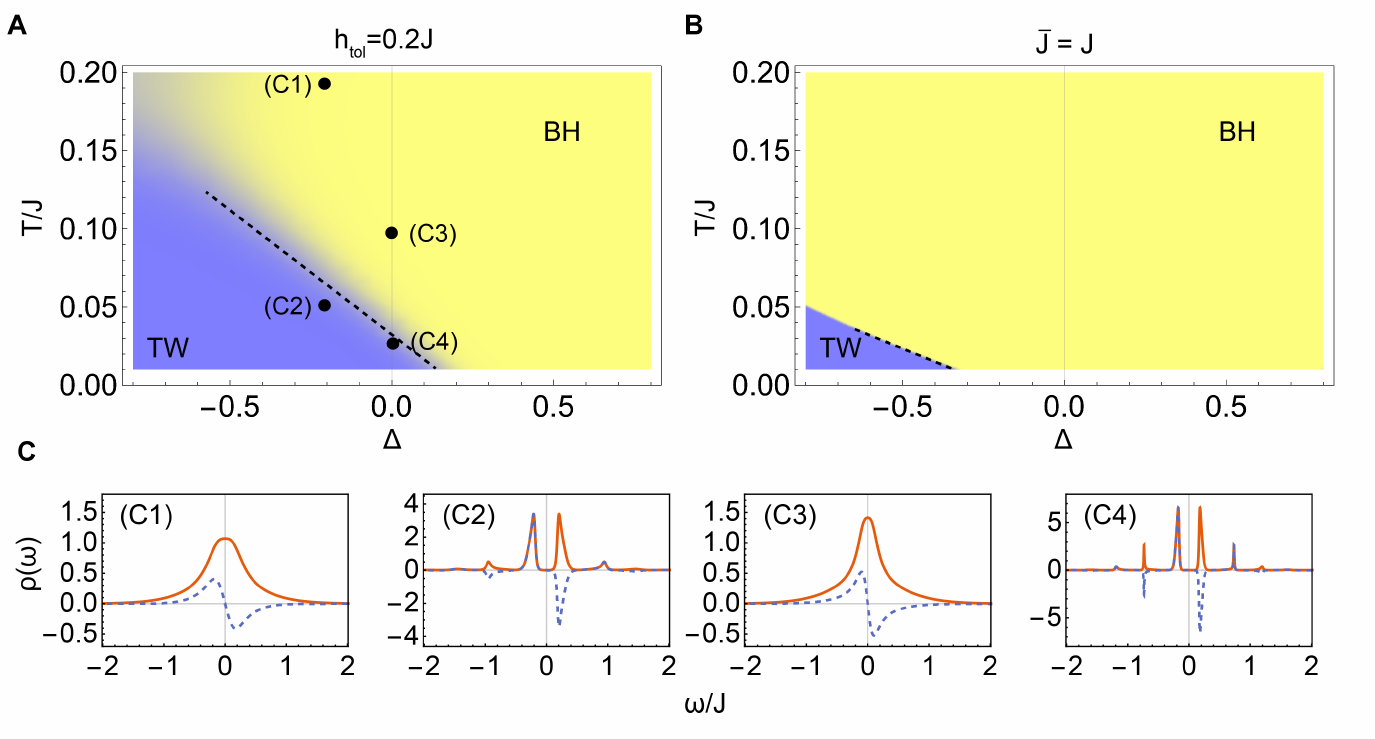}
		\caption{\textbf{Phase Diagrams of the Hamiltonian before and after Quench:} (A) Phase diagram for Hamiltonian before quench. Here $h\neq 0$ and $h_\text{tol}$ determines the spectral functions, and we set $h_\text{tol}=0.2J$. (B) Phase diagram for Hamiltonian after a quench. Here $h=0$ and $\bar{J}$ is the only parameter left, and we set ${\bar J}=J$. $J$ is taken as the energy unit. ``BH" denotes the black hole phase and ``TW" denotes the traversable wormhole phase. They are distinguished by the spectral functions, as illustrated in Fig. \ref{illustration}. The dashed line represents a first-order transition separating two phases, associated with discontinuity in observables, and it is a crossover between two phases for other places. (C) Several typical examples of the spectral functions of BH and TW phases are plotted, with their corresponding locations marked by solid circles in (A).  } 
		\label{phase_diagram}
	\end{figure*}
	
	Knowing the Green's function $G^<_{aa^\prime}(t)$, one can obtain $G^<_{aa^\prime}(\omega)$ by performing the Fourier transformation, and then derive the spectral function $\rho_{aa^\prime}(\omega)=G^{<}_{aa^\prime}(\omega)/2\pi i n_F(\omega)$ by the fluctuation-dissipation theorem, where $n_F(\omega)$ is the Fermi-Dirac distribution. In the BH phase, $\rho_{aa^\prime}(\omega)$ shows a single broad peak near $\omega\sim0$ for $a=a^\prime$, and is significantly smaller for $a\neq a^\prime$. In the TW phase, $\rho_{aa^\prime}(\omega)$ displays multiple narrow peaks. $\rho_{aa^\prime}(\omega)$ behaves similarly for $a=a^\prime$ and $a\neq a^\prime$, except for their different parities in $\omega$. This difference is also shown in Fig. \ref{illustration}(B) and (C). Hence, we have established correspondence between the behavior of the spectral function and the presence of a traversable wormhole. Equipped with this correspondence, we are now ready to move on to discuss the random spin models.

	\vspace{0.05in}
	
\textbf{Connection between the Random Spin Models and the Maldacena-Qi Model.} 
When $\Delta=1$, $h=0$ and $\bar{J}=0$, the model Eq. \eqref{random_spin} possesses $SU(2)$ symmetry, which becomes the Sachdev-Ye model when the $SU(2)$ group is promoted to $SU(M)$ group. The Sachdev-Ye model can be solved in the large-$N$ and large-$M$ limits where various simplifications can be found. Here we will take similar large-$N$ and large-$M$ approximations for studying the quench dynamics of the model Eq. \eqref{random_spin} in general, and the approximations can be justified by qualitative agreements with exact diagonalizations (see Supplementary Materials) and experiments. 	Before studying the quench dynamics, we first address the connection between this model and the Maldacena-Qi model.

	\begin{figure*}[tb]
		\centering
		\includegraphics[width=1.0\linewidth]{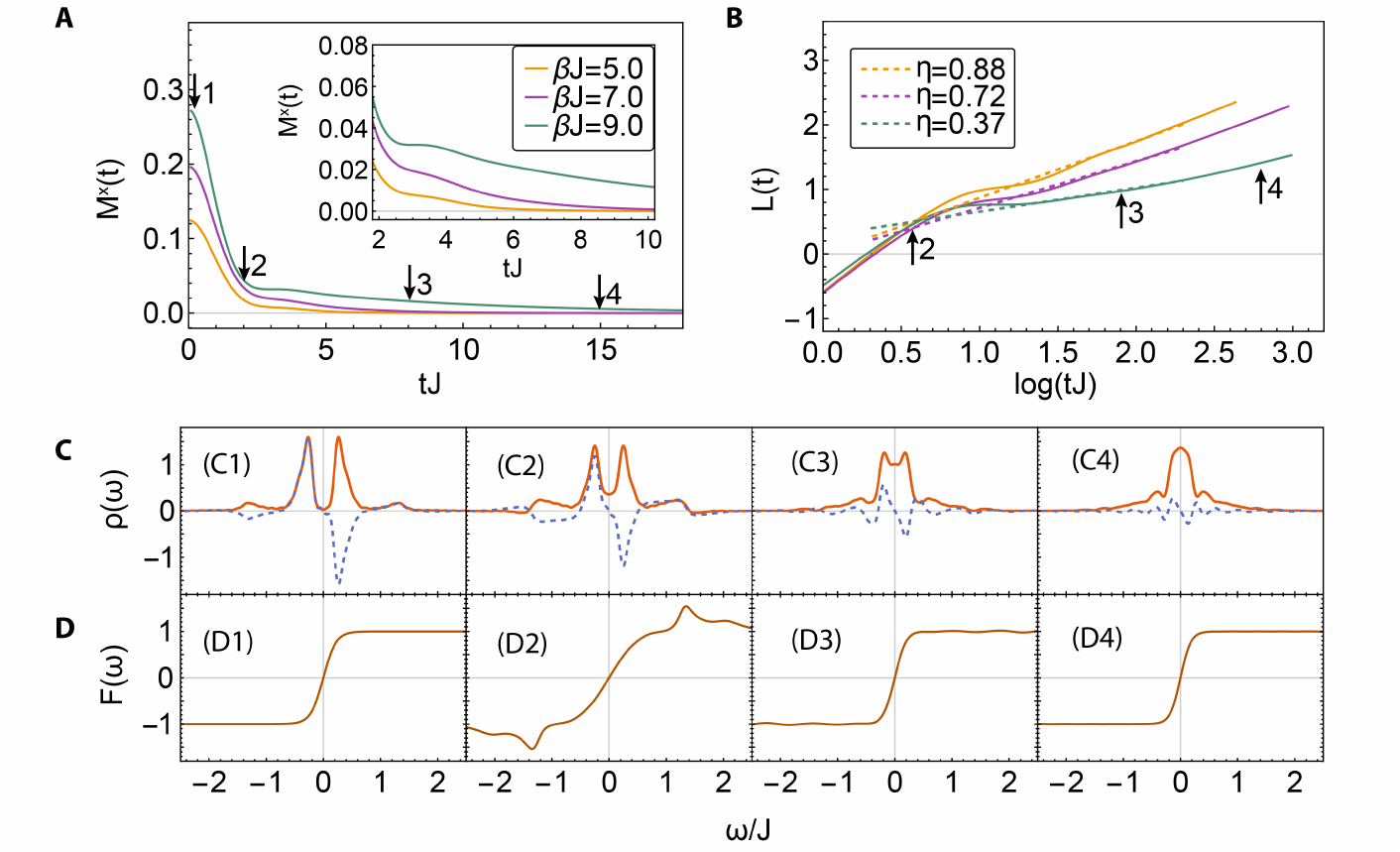}
		\caption{\textbf{Slow Dynamics after Quench:} (A) The decay of the total magnetization $M^x$ along $\hat{x}$ as a function of time $t$ (in unit of $1/J$) for three different temperatures $\beta J$. The inset shows a zoom-in plot for $tJ>t^*J\sim 2$. (B) The same curves as (A), but the loglog-log plot is used. Here the vertical axes $\mathcal{L}(t)$ denotes $\log(|\log(M^x(t)/M^x(0))|)$. The slope of $\mathcal{L}-\log(tJ)$ curves give the exponent in the stretched exponentials. (C) Four representative spectral functions $\rho(\omega)$ and distribution functions $\mathcal{F}(\omega)$ for different stages of the evolution, as marked in (A) and (B). Here we take $\Delta=-0.73$, which is realized in the Rydberg atom experiment \cite{Heidelberg19}, and we set ${\bar J}=J$.   } 
		\label{Dynamics_1}
	\end{figure*}
	
	First of all, the random spin model can be written in terms of fermion operators $\hat{c}_{i,s}$ with spins $s=\uparrow,\downarrow$, by rewriting the spin operators as $\hat{S}^\alpha_i=\frac{1}{2}\hat{c}^\dagger_{i,s}(\sigma^\alpha)_{s s^\prime} \hat{c}_{i,s^\prime}$, where $\alpha=x,y,z$ and $\sigma^\alpha$ denotes the corresponding Pauli matrices. Then, we define the imaginary-time Green's function of fermions $G(\tau)$ as a $2\times 2$ matrix, with the matrix elements defined as 
	\begin{equation}\label{defG}
	G_{s s^\prime}(\tau)=\frac{1}{N}\sum\limits_{i}\left<\mathcal{T}_\tau \hat{c}_{i,s}(\tau)\hat{c}^{\dagger}_{i,s^\prime}(0)\right>.
	\end{equation}
	We first consider the case with $\bar{J}=0$. The Fourier transformation of $G(\tau)$ defines $G(i\omega_n)$, and by the Schwinger-Dyson equation, we have 
	\begin{equation}\label{eq:SD_RD}
	G(i\omega_n)^{-1}=-i\omega_n-\frac{h}{2}\sigma^x-\Sigma(i\omega_n),
	\end{equation}
	with the Matsubara frequency $\omega_n=(2n+1)\pi/\beta$ and $\beta=1/(k_\text{B}T)$ being the inverse temperature, where $\Sigma(i\omega_n)$ is the self-energy in the Matsubara frequency domain. With the large-$N$ and the melon diagram approximations, the self-energy can be obtained as (see Methods for details)
	\begin{equation} 
	\Sigma(\tau)=-\frac{J^2}{4}\sum_{\alpha\alpha^\prime}\xi^\alpha \xi^{\alpha^\prime}\sigma^{\alpha^\prime} G(\tau)\sigma^\alpha\text{Tr}\left[\sigma^{\alpha^\prime} G(\tau)\sigma^\alpha G(-\tau)\right], \label{eq:SelfE}
	\end{equation}
	where $\xi=(1,1,\Delta)$.
	
	For the Maldacena-Qi model, we can similarly define a $2\times 2$ matrix $G(\tau)$ with $G_{aa^\prime}(\tau)= \sum_i \langle \mathcal{T}_\tau \hat{\psi}^a_{i}(\tau)\hat{\psi}^{a^\prime}_{i}(0)\rangle/N$, with $a,a^\prime=\text{L},\text{R}$, and $G(i\omega_n)$ in the Matsubara frequency domain also obeys the Schwinger-Dyson equation as 
	\begin{equation}
	G(i\omega_n)^{-1}=-i\omega_n-\mu\sigma^y-\Sigma(i\omega_n).\label{eq:SD_MQ}
	\end{equation}
	It has been shown that the Green's function obeys the self-consistency equation \cite{MQ}
	\begin{equation}
	\Sigma_{a a^\prime}(\tau)=J^2 G_{a a^\prime}(\tau)^3.\label{eq:selfE_MQ}
	\end{equation}
	
	We first consider the case $\Delta=0$, where direct evaluation of Eq. \eqref{eq:SelfE} yields 
	\begin{equation}\label{eq:selfE_RD}
	\begin{aligned}
	\Sigma_{ss}(\tau)&=-J^2G_{ss}(\tau)G_{\bar{s}\bar{s}}(\tau)G_{\bar{s}\bar{s}}(-\tau),\\
	\Sigma_{s\bar{s}}(\tau)&=-J^2G_{s\bar{s}}(\tau)G_{\bar{s}s}(\tau)G_{s\bar{s}}(-\tau),
	\end{aligned}
	\end{equation}
	where $\bar{s}\neq s$. First of all, to compare Eq. \eqref{eq:SD_RD} with Eq. \eqref{eq:SD_MQ}, we notice that the spin model has rotational symmetry along $\hat{z}$, and therefore one can replace $\sigma^x$ in Eq. \eqref{eq:SD_RD} by $\sigma^y$. Then note that the system is invariant when rotating $\pi$ along $\hat{y}$: $\hat{c}_{i}\rightarrow i\sigma_y \hat{c}_{i}$, which gives $G_{\uparrow \uparrow}(\tau)=G_{\downarrow \downarrow}(\tau)$ and $G_{\uparrow \downarrow}(\tau)=-G_{\downarrow \uparrow}(\tau)$, and the system also has the particle-hole symmetry $\hat{c}_i \rightarrow \hat{c}_i^\dagger$ that corresponds to $(\hat{S}^x_i,\hat{S}^y_i,\hat{S}^z_i)\rightarrow (-\hat{S}^x_i,\hat{S}^y_i,-\hat{S}^z_i)$, which gives $G_{\uparrow \uparrow}(\tau)=-G_{\uparrow \uparrow}(-\tau)$ and $G_{\uparrow \downarrow}(\tau)=-G_{ \downarrow \uparrow}(-\tau)$.
	These symmetry relations simplify Eq. \eqref{eq:selfE_RD} into $\Sigma_{ss'}(\tau)=J^2G_{ss'}(\tau)^3$, which takes the same form as Eq. \eqref{eq:selfE_MQ}. In this way, Eq. \eqref{eq:SD_RD} and Eq. \eqref{eq:selfE_RD} for the random spin model become completely identical to Eq. \eqref{eq:SD_MQ} and Eq. \eqref{eq:selfE_MQ} for the Maldacena-Qi model. Two spin components $s,s^\prime=\uparrow,\downarrow$ play the role of $a,a^\prime=\text{L},\text{R}$, and the Zeeman field $h$ plays the role of the coupling term $\mu$, which is responsible for the wormhole being traversable. The equivalence between these two models also holds approximately for non-zero but small $\Delta$, where a non-zero small $\Delta$ provides an extra channel to couple two sides, as one can see from the low energy effective action (see Supplementary Materials for details). 
	
    Next, we discuss how to deal with a non-zero ${\bar J}$. Up to a constant, the term associated with ${\bar J}$ can be cast into 
	\begin{equation}
	\frac{\bar{J}}{2N}\sum\limits_{\alpha=x,y,z}\xi^\alpha\left(\sum\limits_{i}\hat{S}^\alpha_i\right)\left(\sum\limits_{i}\hat{S}^\alpha_i\right). \label{barJ}
	\end{equation}
	We implement the molecular field approximation by denoting $M^\alpha=\sum_i\langle \hat{S}^\alpha_i\rangle/N$ and $h^\alpha_\text{eff}= \bar{J}\xi^\alpha M^\alpha$. Then, Eq. \eqref{barJ} can be written as
	\begin{equation}
	\sum\limits_{\alpha=x,y,z}h^\alpha_\text{eff}\sum\limits_{i}\hat{S}^\alpha_i.
	\end{equation}  For equilibrium, $M^\alpha$ needs to be determined self-consistently. For dynamics, $M^\alpha$ evolves in time. In this work, since we initially polarize spins along $\hat{x}$, we focus on the situation only $M^x\neq 0$ and $M^y=M^z=0$. Hence, this term can be combined with the $h$-term by replacing $h$ with $h_\text{tol}=h-h^x_\text{eff} = h - \bar{J} M^x$.
	
	\begin{figure*}[tb]
		\centering
		\includegraphics[width=1.0\linewidth]{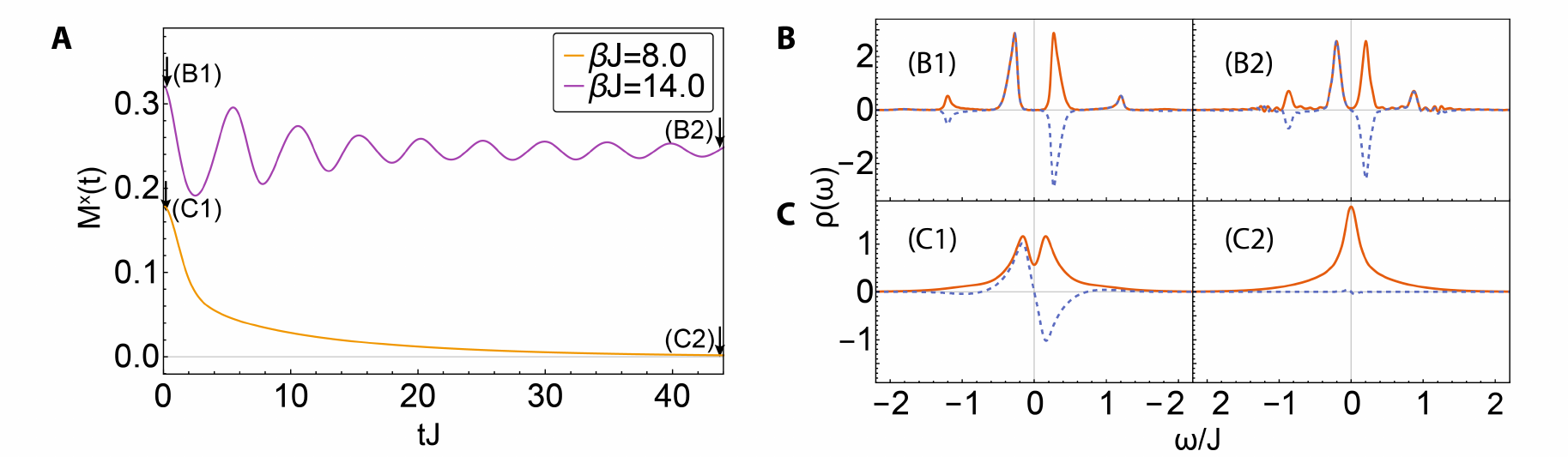}
		\caption{\textbf{Oscillatory Behavior after Quench:} (A) The decay of the total magnetization $M^x$ along $\hat{x}$ as a function of time $t$ (in the unit of $1/J$) for two different temperatures $\beta J$. (B) The spectral functions $\rho(\omega)$ at the initial time of these two cases ((B1) and (C1)), and at the long-time of these two cases ((B2) and (C2)). Here we have taken $\Delta=-0.5$ and ${\bar J}=0$.    } 
		\label{Dynamics_2}
	\end{figure*}
	
	\vspace{0.05in}
	
	\textbf{Phase Diagram.} The quenching process we will consider is instantaneously turning off $h$ from non-zero to zero, therefore, here we will first discuss the equilibrium phase diagram for both finite and zero $h$, respectively. The model parameters contain $J$, $\Delta$, $h$ and ${\bar J}$, as well as temperature $T$. Here we take $J$ as the energy unit. When $h$ is finite, it polarizes spins, and $M^x$ is non-zero, which further normalizes $h_\text{tol}$ through the effect of ${\bar J}$. When the self-consistency is reached, there is a one-to-one correspondence between $h$ and $h_\text{tol}$, and only $h_\text{tol}$ enters the equation that determines the spectral functions. Hence, in Fig. \ref{phase_diagram}(A), we fix $h_\text{tol}$ and plot the phase diagram in terms of the other two dimensionless parameters $T/J$ and $\Delta$. The phase diagram is obtained by numerically solving the self-consistency equations discussed above. Here the ``BH" and ``TW" are distinguished by their different behaviors of spectral functions, as we illustrate in Fig. \ref{illustration} (B) and (C) and show in Fig. \ref{phase_diagram}(C). Consistent with the Maldacena-Qi model \cite{MQ}, the random spin model with small $\Delta$ displays the BH phase at high temperature and the TW phase at low temperature (see Supplementary Material for the discussion of the phase transition). 
	
	When $h=0$, the system possesses a spin rotation symmetry of $(\hat{S}^x_i,\hat{S}^y_i,\hat{S}^z_i)\rightarrow (-\hat{S}^x_i,-\hat{S}^y_i,\hat{S}^z_i)$. If this symmetry is respected, then $M^x$ is always zero and $h_\text{tol}$ is also zero. Nevertheless, the ground state can have non-zero $M^x$ due to the spontaneous symmetry breaking, which leads to non-zero $h_\text{tol}$. Both a non-zero ${\bar J}$ and a non-zero $\Delta$ can lead the system into a TW phase \cite{coupledSYK10SymBreak,coupledSYK11SymBreak}. However, as shown in Fig. \ref{phase_diagram}(B), the TW phase is significantly smaller compared with the $h\neq 0$ case shown in Fig. \ref{phase_diagram}(A).

	\vspace{0.05in}
	
	\textbf{Quench Dynamics.} Now we discuss the two universal features observed in the quench experiments. As we have seen, the $h$-term is mapped to the $\mu$-term in the Maldacena-Qi model, which couples two boundaries and makes the wormhole traversable. Hence, in the gravity picture, the quench dynamics of suddenly turning off $h$ corresponds to turning off the coupling term, which is responsible for making a wormhole traversable. Below we will calculate the quench dynamics numerically by the Kadanoff-Baym formula, which gives how the spectral function evolves in time (see Method). By the correspondence between the spectral function and the space-time geometry discussed above, we can establish a gravity picture of the quench dynamics.   
	
	Firstly, we discuss the slow dynamics described by a stretched exponential. Initially, the spectral function is a typical TW-type one, as shown in Fig. \ref{Dynamics_1}(C1). We show in Fig. \ref{Dynamics_1}(A) that the dynamics contain two stages. In the first stage when $tJ< t^*J$, with $t^*J\approx 2$ in the case shown in Fig. \ref{Dynamics_1}(A), $M^x$ decays quite fast. As shown in Fig. \ref{Dynamics_1}(C2), the spectral function still retains TW-phase-like in this stage. It is known that the distribution function $\mathcal{F}(\omega)$ at equilibrium should be $\tanh \left(\beta \omega/2\right)$, as in the initial case shown in Fig. \ref{Dynamics_1}(D1). As one can see in Fig. \ref{Dynamics_1}(D2), at the end of the first stage, the system strongly deviates from equilibrium.

	In the second stage, with $tJ$ ranging from $t^*J$ to a few tens, the dynamics is much slower. Note that if $M^x$ obeys a stretched exponential $e^{-C(t/t_0)^\eta}$, it should manifest as a straight line in the $\log(|\log M^x|)-\log t$ plot, and the slope determines $\eta$. We show this in Fig. \ref{Dynamics_1}(B), and the fitting yields $\eta<1$, meaning that the dynamics is slower than usual exponential decay. In Fig. \ref{Dynamics_1}(C3), we show that the multiple peaks in the spectral function gradually merge into a single peak, and meanwhile, $\rho_{ss^\prime}$ with $s\neq s^\prime$ is gradually suppressed. At the long-time limit, as shown in Fig. \ref{Dynamics_1}(C4), the spectral function becomes BH-phase-like, where $\rho_{ss^\prime}$ shows a single broad peak around $\omega\sim 0$ for $s=s^\prime$ and is vanishing small for $s\neq s^\prime$. Hence, in the gravity picture, this process corresponds to the wormhole gradually closing. From Fig. \ref{Dynamics_1}(D3),(D4), we also see that the distribution functions fast reach quasi-equilibrium in this stage, and by fitting the distribution function with $\tanh (\tilde{\beta} \omega/2)$, we can obtain an effective temperature $1/\tilde{\beta}$ for the saturated state. By comparing Fig. \ref{Dynamics_1}(D4) with (D1), one can see that the temperature for the final state is close to the temperature of the initial state. By examining various different parameters (see Supplementary Material for more examples), we identify that the stretched exponential decay occurs generically when a traversable wormhole is disconnected and becomes two decoupled black holes, and the temperatures of the initial TW and the final BH phases are close.  
	
	Secondly, we discuss the oscillatory behavior. In Fig. \ref{Dynamics_2} we compare two cases which only differ by the parameter $\beta J$ but show two different behaviors. For the one with a smaller $\beta J$, $M^x$ monotonically decays to zero as discussed above. For the one with a larger $\beta J$, $M^x$ saturates to a finite value in the long-time limit, and oscillates in time before saturation. Here we also want to find a generic condition to determine which behavior should take place, and it turns out that the answer is also quite clear in the gravity picture. As we show in Fig. \ref{Dynamics_2}(B) and (C), these two cases have similar initial states in the TW phase. However, their long-time saturation states are very different. $\rho(\omega)$ at long-time exhibits typical behavior of the BH phase for the one with smaller $\beta J$, but exhibits typical behavior of the TW phase for the one with larger $\beta J$. Note that in Fig. \ref{phase_diagram}(B), even with $h=0$, there still exists TW phase in the equilibrium phase diagram, which means that turning off $h$ does not always disconnect the wormhole. If the long-time saturation phase is still retained in the TW phase, there will exhibit an oscillatory behavior of magnetization in time. This oscillatory behavior can also be found by using the effective action with gravity dual (see Supplementary Material). By examining different parameters (see Supplementary Material), we also find that whether the monotonical decay or oscillatory behavior occurs in the quench dynamics essentially depends on whether the final state is the BH phase or the TW phase.           
	
	\vspace{0.05in}
	
	\textbf{Summary and Outlook.} In summary, we show that the quench dynamics of randomly interacting quantum spins after turning off an external field can be understood in the dual gravity picture as turning off a coupling field for making the wormhole traversable. If this process disconnects the traversable wormhole and finally yields two decoupled black holes, the magnetization displays a slow decay described by the stretched exponential function toward zero magnetization. If this process does not disconnect the traversable wormhole, due to the existence of the residual couplings between the boundary fields, the magnetization saturates to a non-zero value at long time and oscillates around the saturation value. This gravity picture conceptually unifies these two universal phenomena observed in the random spin model realized in different physical systems.    
	
	The stratagem of our study is reminiscent of the concept of the``fixed point" in the renormalization group theory, where different microscopic models can share the same low-energy description given by a fixed point action. Here we use the Maldacena-Qi model, whose gravity dual has been shown explicitly, as the ``holographic fixed point", and we use the holographic fixed point to define correspondence between quantum properties and the gravity properties. We further share this correspondence with different microscopic models in the neighborhood of the Maldacena-Qi model with similar quantum properties, and these models are much closer to reality. Therefore, the success of this approach can inspire more applications on realistic quantum models with insights from the gravity theory.        
	
	\vspace{0.05in}
	
	\textit{Acknowledgment.} We thank Yiming Chen, Bartek Czech, Ruihua Fan, Yingfei Gu, Chao-Ming Jian, Xiao-Liang Qi and Yi-Zhuang You for helpful discussions. This work is supported by Beijing Outstanding Young Scientist Program (HZ), NSFC Grant No. 11734010 (HZ and YC), NSFC under Grant No. 11604225 (YC), MOST under Grant No. 2016YFA0301600 (HZ) and Beijing Natural Science Foundation (Z180013) (YC).
	
	\vspace{0.05in}
	
	\textbf{Methods.}
	
	\textit{Derivation of the Self-Consistent Equations.} The original model given by Eq. \eqref{random_spin} can be written by fermion representation with $\hat{S}^\alpha_i=\sum_{s s^\prime}\hat{c}^{\dagger}_{i,s}(\sigma^\alpha)_{ss^\prime}\hat{c}_{i,s^\prime}/2$ and the constrain $\sum_s\hat{c}^{\dagger}_{i,s}\hat{c}_{i,s}=1$. We now promote this model by adding an additional $M$ indices as   
	\begin{equation}
	\hat{H}= \frac{1}{\sqrt{M}} \sum_{ij,\alpha \gamma} J_{ij}\xi^\alpha\hat{T}^{\alpha,\gamma}_i\hat{T}^{\alpha,\gamma}_j-h\sum_i\hat{S}^x_i.
	\end{equation}
	where $$\hat{T}^{\alpha,\gamma}_i=\frac{1}{2}\sum_{s_i,m_i}\hat{c}^{\dagger}_{i,s_1,m_1}(\sigma^\alpha)_{s_1s_2}(T^\gamma)_{m_1m_2}\hat{c}_{i,s_2,m_2}.$$ Here $m_i=1,2,...M$ and $T^\gamma$ are the generators of the $SU(M)$ group that satisfies $$\sum_\gamma T^\gamma_{m_1m_2}T^\gamma_{m_3m_4}=\delta_{m_1m_4}\delta_{m_2m_3}-\frac{1}{M}\delta_{m_1m_2}\delta_{m_3m_4}.$$ The external field $h$ only couples to the $SU(2)$ part. The constrain is also promoted as $\sum_{s,m}\hat{c}^{\dagger}_{i,s,m}\hat{c}_{i,s,m}=M$. Firstly, we take the imaginary time approach in the large-$N$ and large-$M$ limits. The constrain is satisfied automatically due to the particle-hole symmetry, and the self-energy can be obtained by the melon diagrams as \cite{SY}
	\begin{equation}
	\begin{aligned} 
	\Sigma^{(M)}(\tau)=-&\frac{J^2}{4}\sum_{\alpha\alpha^\prime}\xi^\alpha \xi^{\alpha^\prime}\sigma^{\alpha^\prime} G^{(M)}(\tau)\sigma^\alpha\\
	&\times\text{Tr}\left[\sigma^{\alpha^\prime} G^{(M)}(\tau)\sigma^\alpha G^{(M)}(-\tau)\right]. \label{eq:SelfEM}
	\end{aligned}
	\end{equation}
	Here we have defined $$
	G^{(M)}_{s s^\prime}(\tau)=\frac{1}{NM}\sum\limits_{i,m}\left<\mathcal{T}_\tau \hat{c}_{i,s,m}(\tau)\hat{c}^{\dagger}_{i,s^\prime,m}(0)\right>.$$ By taking $M\rightarrow 1$, we obtain Eq. \eqref{defG} and Eq. \eqref{eq:SelfE} in the maintext with $\Sigma(\tau)=\Sigma^{(1)}(\tau)$ and $G(\tau)=G^{(1)}(\tau)$.

	Secondly, we also employ the real-time approach and we consider the $M=1$ case. We begin with the greater and lesser Green's functions in real-time defined as
	\begin{equation}\label{eq:M_GF_keldysh}
	\begin{aligned}
	G^>_{ss'}(t_1,t_2)&\equiv-i  \frac{1}{N} \sum_{l} \left<c_{l,s}(t_1)c_{l,s'}^\dagger(t_2)\right>,\\
	G^<_{ss'}(t_1,t_2)&\equiv i  \frac{1}{N} \sum_{l} \left<c_{l,s'}^\dagger(t_2)c_{l,s}(t_1)\right>,
	\end{aligned}
	\end{equation}
	as well as the retarded and advanced Green's function defined as
	\begin{equation}
	G^{R/A}_{ss'}(t_1,t_2)\equiv \mp i\Theta\left(\pm t_{12}\right) \frac{1}{N} \sum_{l}  \left<\{c_{l,s}(t_1),c_{l,s'}^\dagger(t_2)\}\right>,
	\end{equation}
	where $\Theta\left( t\right)$ is the Heaviside step function and we have defined $t_{12}=t_1-t_2$. $G^{R/A}$ is related to $G^{\gtrless}$ as 
	\begin{equation}\label{relationGRA}
	G^{R/A}(t_1,t_2)=\pm \Theta\left(\pm t_{12}\right)\left(G^>(t_1,t_2)-G^<(t_1,t_2)\right).
	\end{equation}
	In the thermal equilibrium, all Green's functions are only functions of $t_{12}$ due to the time-translational symmetry. To solve the real-time Green's functions self-consistently, we introduce the spectral function as 
	\begin{equation}\label{eq:GR_rho}
	G^{R}(\omega) = \int \diff z  \frac{\rho(z)}{z - \omega + \iu 0}, 
	\end{equation}
	which implies $\rho(\omega)=-\text{Im}G^R(\omega)/\pi$. Other Green's functions are determined by the fluctuation-dissipation theorem as
	\begin{equation}
	\begin{aligned}
	G^<(\omega)&=2\pi i n_F(\omega)\rho(\omega),\\
	G^>(\omega)&=-2\pi i n_F(-\omega)\rho(\omega),
	\end{aligned}
	\end{equation}
	where $n_F(\omega)$ is the Fermi-Dirac distribution function.
	
	To derive the self-consistent equation for the spectral function, we apply the Fourier transformation on the Eq.~\eqref{eq:SelfE} $\Sigma(\tau) \to \Sigma(\iu \omega_n)$ and perform the analytical continuation $\iu \omega_n \to \omega + \iu 0$ after the Matsubara frequency summation for the internal dummy frequency \cite{K2}. We then obtain $G^R(\omega)^{-1}=\omega+\frac{h}{2}\sigma^x-\Sigma^{R}(\omega)$, with
	\begin{equation}\label{eq:SelfSigmaR}
	\begin{split}
	\Sigma^R(\omega) &=  \frac{J^2}{4} \int_{-\infty}^{\infty} \diff t \  \Theta(t) e^{\iu \omega t}   \\
	& \sum_{\alpha \alpha'} \Bigg( \xi^{\alpha} \xi^{\alpha'} \sigma^{\alpha'} G^>(t) \sigma^{\alpha} \Tr \left[\sigma^{\alpha} G^<(-t) \sigma^{\alpha'} G^>(t) \right] - \\
	& \qquad \xi^{\alpha} \xi^{\alpha'} \sigma^{\alpha'} G^<(t) \sigma^{\alpha} \Tr \left[\sigma^{\alpha} G^>(-t) \sigma^{\alpha'} G^<(t) \right] 
	\Bigg). \\
	\end{split}
	\end{equation}
	Iteratively solving these equations gives the spectral function $\rho(\omega)$ and real-time Green's functions.
	
	\vspace{0.05in}
	
	\textit{The Kadanoff-Baym Equation.} To study the real-time dynamics, we utilize the Kadanoff-Baym equation on the Keldysh contour \cite{kamenev2011field}, which describes the real-time evolution of $G^{\gtrless}$. Assuming the melon diagram approximation Eq. \eqref{eq:SelfE} and applying the Langreth rules \cite{stefanucci2013nonequilibrium} on the Schwinger-Dyson equation, we find that \cite{quench}: 
	\begin{equation}
	\begin{aligned}\label{KBeq}
	i\partial_{t_1}&G^\gtrless(t_1,t_2)+ \frac{h_{\text{eff}}(t_1)\sigma^x}{2}G^\gtrless(t_1,t_2)=\\ &\int d t_3 (\Sigma^R(t_1,t_3)G^\gtrless(t_3,t_2)+\Sigma^\gtrless(t_1,t_3)G^A(t_3,t_2)), \\
	-i\partial_{t_2}&G^\gtrless(t_1,t_2)+G^\gtrless(t_1,t_2)\frac{h_{\text{eff}}(t_2)\sigma^x}{2}=\\ &\int d t_3 (G^R(t_1,t_3)\Sigma^\gtrless(t_3,t_2)+G^\gtrless(t_1,t_3)\Sigma^A(t_3,t_2)).
	\end{aligned}
	\end{equation}
	Here we have taken the contribution of $\bar{J}$ into account by 
	\begin{equation} \label{heff}
	\begin{aligned}
	h^x_{\text{eff}}(t)=i\frac{\bar{J}}{2}\left(G^<_{\uparrow\downarrow}(t,t)+G^<_{\downarrow\uparrow}(t,t)\right).
	\end{aligned}
	\end{equation}
	The real-time self-energies in Eq. \eqref{KBeq} are given by 
	\begin{equation}\label{selfreal}
	\begin{aligned}
	\Sigma^{\gtrless}&(t_1,t_2)=\frac{J^2}{4}\sum_{\alpha,\alpha^\prime}\xi^\alpha \xi^{\alpha^\prime} \\&\sigma^{\alpha^\prime} G^{\gtrless}(t_1,t_2)\sigma^\alpha\text{tr}\left[\sigma^{\alpha^\prime} G^{\gtrless}(t_1,t_2)\sigma^\alpha G^{\lessgtr}(t_2,t_1)\right],
	\end{aligned}
	\end{equation}
	with $\Sigma^{R/A}$ related to $\Sigma^{\gtrless}$ as
	\begin{equation}\label{relationSRA}
	\Sigma^{R/A}(t_1,t_2)=\pm \Theta\left(\pm t_{12}\right)\left(\Sigma^>(t_1,t_2)-\Sigma^<(t_1,t_2)\right).
	\end{equation}
	For $t_1,t_2<0 $, $G^\gtrless_\chi(t_1,t_2)=G^\gtrless_\chi(t_{12})$ is given by the equilibrium solution, which serves as the initial conditions for the time dynamics.  Evolving $G^\gtrless_\chi(t_1,t_2)$ using Eq. \eqref{KBeq}-\eqref{selfreal} gives the quantum dynamics. 
	
	To analyze the quench dynamics, we study the spectral function and the quantum distribution function at given time $t$. We firstly separate the relative time $t_r$ by defining the temporal Green's function $\tilde{G}^\iota(t,t_r)$ as
	\begin{equation}\label{tilde}
	\begin{aligned}
	&\tilde{G}^\iota(\text{max}\{t_1,t_2\},t_1-t_2)=G^\iota(t_1,t_2),\\
	&\tilde{G}^\iota(t,\omega)=\int dt_r e^{i\omega t_r}\tilde{G}^\iota(t,t_r).
	\end{aligned}
	\end{equation} 
	Here $\iota\in \{>,<, R, A\}$ and we perform the Fourier transformation for the relative time. The temperal spectral function is then defined as
	\begin{equation}
	\rho(t,\omega)=-\frac{1}{\pi}\text{Im} \tilde{G}^R(t,\omega).
	\end{equation}
	Similar as Eq. \eqref{tilde}, we could also define $\tilde{\Sigma}^\iota$. At thermal-equilibrium, there is no time dependence for both $\tilde{G}$ and $\tilde{\Sigma}$, and the fluctuation-dissipation theorem implies the relation \cite{kamenev2011field}
	\begin{equation}\label{FDT}
	\tanh\frac{\beta \omega}{2}=\frac{\tilde{\Sigma}^>(\omega)+\tilde{\Sigma}^<(\omega)}{\tilde{\Sigma}^R(\omega)-\tilde{\Sigma}^A(\omega)}.
	\end{equation}
	Then for the non-equilibrium case, motivated by Eq. \eqref{FDT}, we define the quantum distribution function at time $t$ as 
	\begin{equation}
	\mathcal{F}(t,\omega)=\frac{\tilde{\Sigma}^>(t,\omega)+\tilde{\Sigma}^<(t,\omega)}{\tilde{\Sigma}^R(t,\omega)-\tilde{\Sigma}^A(t,\omega)}.
	\end{equation}

\onecolumngrid
\begin{center}
\newpage\textbf{\large
Supplementary Materials: Disconnecting a Traversable Wormhole: \\ Universal Quench Dynamics in Random Spin Models}
\\
\vspace{4mm}

{Tian-Gang Zhou,$^1$ Lei Pan,$^{1}$, Yu Chen,$^{2}$ Pengfei Zhang,$^{3,4,*}$ and Hui Zhai$^{1,\dagger}$}\\
\vspace{2mm}
{\em \small
$^1$Institute for Advanced Study, Tsinghua University, Beijing,100084, China\\
$^2$Graduate School of China Academy of Engineering Physics, Beijing, 100193, China\\
$^{3}$Institute for Quantum Information and Matter,\\ California Institute of Technology, Pasadena, California 91125, USA\\
$^4$Walter Burke Institute for Theoretical Physics, California Institute of Technology, Pasadena, California 91125, USA}
\end{center}

\setcounter{equation}{0}
\setcounter{figure}{0}
\setcounter{table}{0}
\setcounter{section}{0}
\setcounter{page}{1}
\makeatletter
\renewcommand{\thepage}{S\arabic{page}} 
\renewcommand{\thesection}{S\arabic{section}}  
\renewcommand{\thetable}{S\arabic{table}}  
\renewcommand{\thefigure}{S\arabic{figure}}
\renewcommand{\theequation}{S\arabic{equation}}

\section{Exact diagonalization results}

Here we report the results from the exact diagonalization calculation of a finite-size spin system, which shows both the slow dynamics and the oscillatory dynamics in different parameter regimes. The results are consistent with those reported in the main text based on the large-$N$ expansion and assuming the melon diagrams. This consistency provides supports to our approximations used in the main text. Nevertheless, we should also emphasize that, because the finite-size exact diagonalization and large-$N$ theory are two different approximations, we do not expect a quantitative agreement between these two calculations.   

\begin{figure*}[tb]
	\centering
	\includegraphics[width=0.9\linewidth]{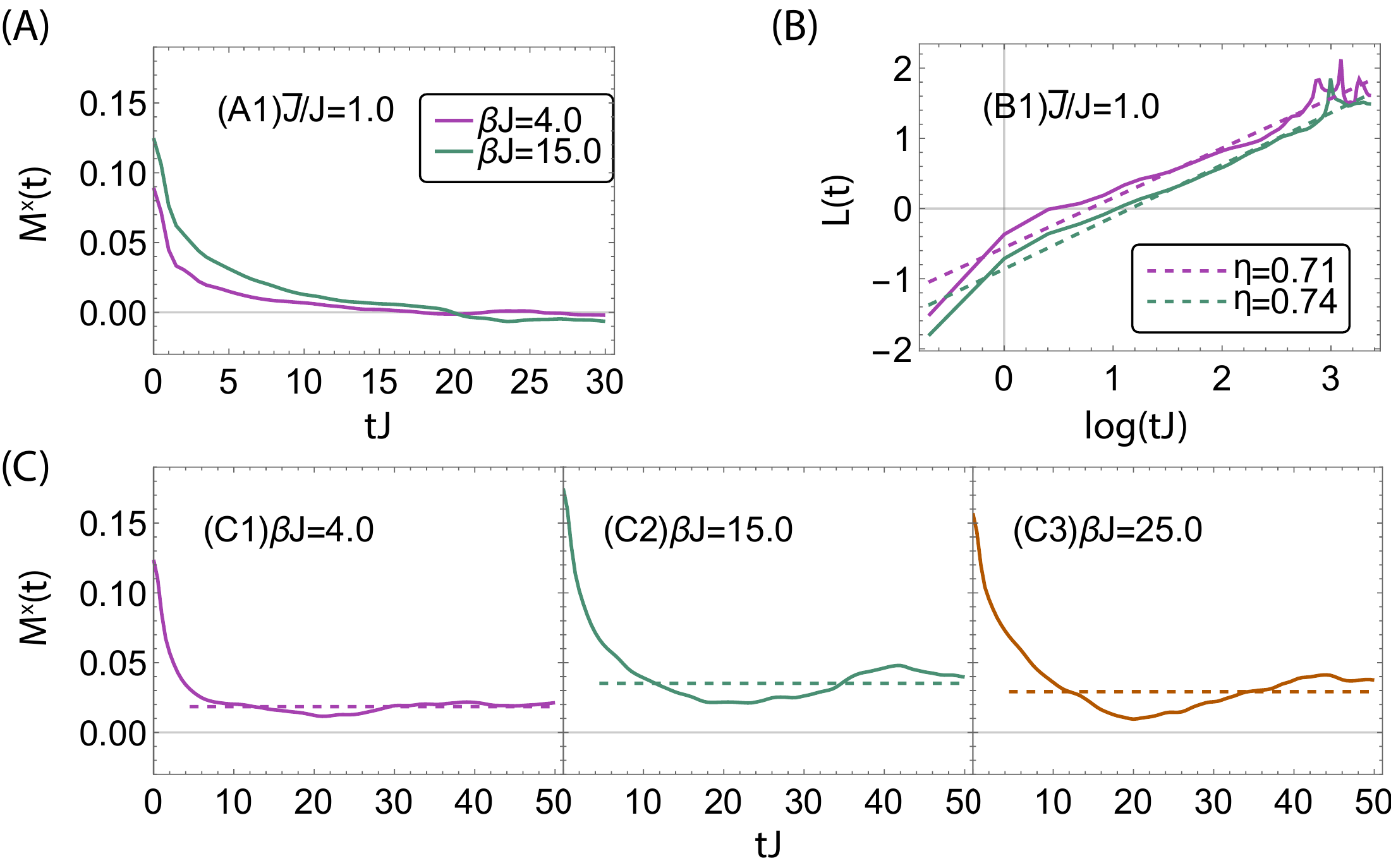}
	\caption{\textbf{Result of Exact Diagonalization.} The results from the exact diagnolization calculation of the random spin model Eq.~(1) in the main text, with the number of spins $N=10$. The quench dynamics is averaged over $60$ different disorder configurations. (A) Quench dynamics in $\bar{J}=0$ and $\Delta=-0.73$, corresponding to the Fig.~3 in the main text. Initially, the system is prepared at $h=0.2 J$, and we completely turn off $h$ to observe the stretched exponential decay. (B) The same data but in the $\log(|\log M^x|)-\log t$ coordinates. Here we show the decay exponent $\eta$ in the figure. (C) Quench dynamics in $\bar{J}=0$ and $\Delta=-0.5$, corresponding to the Fig.~4 in the main text. Initially system is prepared at $h=0.2 J$. For (A) and (B), $h$ is quenched to zero, but for (C), $h$ is quenched to $0.02J$. The dashed line in (C) is the long-time averaged value. }
	\label{fig:S_exactdiagonalization}
\end{figure*}

Firstly, slow dynamics is shown in the Fig.~\ref{fig:S_exactdiagonalization}(A) and (B). In the same parameter region as in our large-$N$ theory discussed in the main text, we observe the stretched exponential decay behavior of $M^x$. In the $\log(|\log M^x|)-\log t$ coordinates, we find the decay exponent $\eta<1$. 

Secondly, we find some hints for the oscillatory behavior as shown in Fig.~\ref{fig:S_exactdiagonalization}(C). Here we should first note that the symmetry breaking behaviors can hardly be achieved in such a small size spin system with $N=10$ by the exact diagonalization method. In fact, we do not find oscillatory behavior in the exact diagonalization calculation when we quench the external field $h$ to zero. This is actually consistent with our conclusion that the oscillatory behavior is associated with the final state in the TW phase, where $M^x$ is non-zero. Hence, to show the oscillatory behavior in the exact diagonalization calculations, we quench $h$ to a very small value, say, $h=0.02$, instead of $h=0$. The results are shown in Fig.~\ref{fig:S_exactdiagonalization}(C). It shows that for small $\beta J$, at a long time the magnetization saturates to a constant, with very small variation. As $\beta J$ increases, the long time dynamics exhibits a larger and larger variation around its long-time averaged value, although the variation is not a perfect oscillation due to the finite-size effect.

\section{Low energy effective action analysis}
\label{sec:lowEAction}

\subsection{Equilibrium solution}

Here we will discuss the low-energy effective action for the random spin model in the TW phase. We can identify two parts of contributions in addition to the Schwarzian action. The first part corresponds to the transverse magnetic field $h_{\text{tol}}$, the effect of which has already been considered in the Maldacena-Qi's work \cite{MQ}. The second part represents the effect from a  non-zero $\Delta$, with the assumption that $| \Delta | \ll 1$. Here we show how these two quantities affect the energy gap $E_{\mathrm{gap}}$ in the TW phase, and the equilibrium magnetization $M^x$.

To begin with, it is known that under the melon-diagram approximation, the effective action in the imaginary time can be written as
\begin{equation}\label{eq:action_origin}
\begin{split}
\frac{\mathcal{I}}{N} & = - \Tr \log  (\partial_{\tau} - \Sigma) 
- \int 
\diff \tau_1 \diff \tau_2
\Bigg[
\Tr \left[  \Sigma (\tau_1, \tau_2) G (\tau_2, \tau_1)  \right]  + \sum_{\alpha \alpha'} \frac{J^2}{16} \xi^{\alpha} \xi^{\alpha'} \left( \Tr \left[ \sigma^{\alpha} G(\tau_1, \tau_2) \sigma^{\alpha'} G(\tau_2, \tau_1) \right] \right)^2
\Bigg]  \\
& \qquad\qquad + \int \diff \tau_1 \frac{h_\text{tol}}{2} \Tr \left[ G(\tau_1, \tau_1) \sigma_x \right]. \\
\end{split}
\end{equation}
Here we define the imaginary-time Green's function as:
\begin{equation}\label{eq:S_newG}
G_{s s^\prime}(\tau)=\frac{1}{N}\sum\limits_{i}\left<\mathcal{T}_\tau \hat{c}_{i,s}(\tau)\hat{c}^{\dagger}_{i,s^\prime}(0)\right>.
\end{equation}
Here $s, s' = \uparrow$ or $\downarrow$ are the up or down spin indices respectively. From now on, we focus on the small $h_{\text{tol}}/J$, where the dynamics can be described by the evolution of reparametrization modes. After imposing the spin rotational symmetries and the particle-hole symmetries discussed in the main text, direct calculation shows 
\begin{equation}\label{eq:S_G4term}
\begin{split}
& \sum_{\alpha \alpha'} \frac{J^2}{16} \xi^{\alpha} \xi^{\alpha'} \left( \Tr \left[ \sigma^{\alpha} G(\tau_1, \tau_2) \sigma^{\alpha'} G(\tau_2, \tau_1) \right] \right)^2 \\
=& 
\frac{J^2}{4}  \Big[ \left(\Delta^2+2\right) G_{\uparrow\uparrow}(\tau_1, \tau_2 ){}^4+\left(\Delta^2+2\right)
G_{\uparrow\downarrow}(\tau_1, \tau_2 ){}^4+2 (\Delta -4) \Delta  G_{\uparrow\downarrow}(\tau_1, \tau_2 ){}^2 G_{\uparrow\uparrow}(\tau_1, \tau_2 ){}^2\Big].
\\
\end{split}
\end{equation}
To the order of $\mathcal{O}(\Delta)$, Eq. \eqref{eq:S_G4term} can be truncated as 
\begin{equation}\label{eq:S_G4term_truncated}
\begin{split}
\frac{J^2}{2}  \Big[ G_{\uparrow\uparrow}(\tau_1, \tau_2 ){}^4 + G_{\uparrow\downarrow}(\tau_1, \tau_2 ){}^4 - 4 \Delta  G_{\uparrow\downarrow}(\tau_1, \tau_2 ){}^2 G_{\uparrow\uparrow}(\tau_1, \tau_2 ){}^2\Big].
\\
\end{split}
\end{equation}
When $\Delta=0$, this effective action is identical to the effective action of the Maldacena-Qi model. After considering the last term in Eq.~\eqref{eq:S_G4term_truncated} and rescaling the time variable, the effective action for the reparametrization modes becomes
\begin{equation}\label{eq:S_lowE_action_full}
\begin{split}
\frac{\mathcal{I}}{N}&= \int d  \tilde{\tau}
\Bigg\{ \left(\left\{\tanh \frac{ t_{l}(\tilde{\tau})}{2}, \tilde{\tau}\right\}+\left\{\tanh \frac{ t_{r}(\tilde{\tau})}{2}, \tilde{\tau}\right\}\right) +
\tilde{h} \left[\frac{t_{l}^{\prime}(\tilde{\tau}) t_{r}^{\prime}(\tilde{\tau})}{\cosh ^{2}  \frac{ \left( t_{l}(\tilde{\tau})-t_{r}(\tilde{\tau}) \right) }{2}}\right]^{\mathcal{D}}  \Bigg\} \\
& \qquad \qquad + \int \diff \tilde{\tau}_1 \diff \tilde{\tau}_2 \tilde{\Delta} \left[\frac{t_{l}^{\prime}(\tilde{\tau}_1) t_{r}^{\prime}(\tilde{\tau}_2)}{\cosh ^{2}  \frac{ \left( t_{l}(\tilde{\tau}_1)-t_{r}(\tilde{\tau}_2) \right) }{2}}\right]^{2\mathcal{D}} 
\left[\frac{t_{l}^{\prime}(\tilde{\tau}_1) t_{l}^{\prime}(\tilde{\tau}_2)}{ \sinh ^{2} \big| \frac{ \left( t_{l}(\tilde{\tau}_1)-t_{l}(\tilde{\tau}_2) \right) }{2}  \big| }\right]^{2\mathcal{D}}. \\
\end{split}
\end{equation}
where $\tilde{\tau} = \frac{J}{\sqrt{2} \alpha_s} \tau$ is the dimensionless imaginary coordinates time. For the SYK$_4$ model, $\alpha_s \approx 0.007$ is a positive numerical factor in the action, and the scaling dimension $\mathcal{D}=1/4$. It's known that the Schwarzian action itself describes the low energy fluctuations of the SYK model, and here we use $\{f, \tau\}$ to denote the Schwarzian derivative. Variable $t_l(\tilde{\tau})$ and $t_r(\tilde{\tau})$ in the low energy action are the time reparametrization modes on the left and right boundaries respectively. We have neglected other modes including the charge fluctuation modes, since they are not excited in our cases. The rescaled coupling constants are
\begin{equation}\label{eq:S_action_para}
\begin{split}
\tilde{h} &= \frac{ h_{\text{tol}} }{4 J} \left[ \frac{2  \alpha_s^2}{\pi}\right]^{1/4}, \qquad\qquad \tilde{\Delta} =  \frac{\Delta}{8 \pi}. \\
\end{split}
\end{equation}
These two parameters are related to the magnetic field $h_{\text{tol}}$ and the anisotropic paramter $\Delta$. In the equilibrium case, the classical solution assumes the ansatz $t_l(\tilde{\tau})=t_l(\tilde{\tau})=t' \tilde{\tau}$, where $t'$ is a constant. The $\tilde{\Delta}$ term could be simplified as:
\begin{equation} \label{eq:S_action_equilibrium_Delta}
\begin{split}
(\Delta \text{ term})=&\int_0^{\tilde{\beta}} \diff \tilde{\tau}_1 \int_0^{\tilde{\beta}} \diff \tilde{\tau}_2 \tilde{\Delta} \left[\frac{t_{l}^{\prime}(\tilde{\tau}_1)  t_{r}^{\prime}(\tilde{\tau}_2)}{\cosh ^{2}  \frac{ \left( t_{l}(\tilde{\tau}_1)-t_{r}(\tilde{\tau}_2) \right) }{2}}\right]^{2\mathcal{D}} 
\left[\frac{t_{l}^{\prime}(\tilde{\tau}_1) t_{l}^{\prime}(\tilde{\tau}_2)}{ \sinh ^{2} \big| \frac{ \left( t_{l}(\tilde{\tau}_1)-t_{l}(\tilde{\tau}_2) \right) }{2}  \big| }\right]^{2\mathcal{D}}  \\
=&\int_0^{\tilde{\beta}} \diff x_1 \int_{-\tilde{\beta}/2}^{\tilde{\beta}/2} \diff x_2
\frac{2 t'{}^2 \tilde{\Delta}}{\sinh \left| t' x_2 \right|}    \\
=&\int_{0}^{\tilde{\beta}} \diff x_1 \int_{\Lambda}^{\infty} \diff x_2 
\frac{4 t'{}^2 \tilde{\Delta}}{\sinh  (t' x_2) }.    \\
\end{split}
\end{equation}
We have employed the ansatz and introduced the new coordinates $x_1 = \frac{1}{2} (\tilde{\tau}_1 + \tilde{\tau}_2),\ x_2 = (\tilde{\tau}_1 - \tilde{\tau}_2)$. The dimensionless value $\tilde{\beta} = \frac{\beta J}{\sqrt{2} \alpha_s}$. Since the UV divergence of the conformal solution is not physical, we have introduced a natural cutoff $\Lambda=(\epsilon/J) \frac{J}{\sqrt{2} \alpha_s}$, where $\epsilon$ is a constant of the order $\mathcal{O}(1)$. Besides, because the effective action describes a zero temperature wormhole, the upper bound of the integration $x_2$ extends to infinity. 

Adding up all the terms, the action reads
\begin{equation}\label{eq:S_action_equilibrium_full}
\begin{split}
\mathcal{I} / \beta = - \frac{J t'{}^2}{\sqrt{2} \alpha_s}
+ \frac{h_{\text{tol}}}{4 (2\pi \alpha_s^2)^{1/4} } \sqrt{t'}
+ \frac{ \Delta  J t' \log \left(\tanh \left(\frac{t' \epsilon }{2
		\sqrt{2} \alpha _s}\right)\right)}{2 \sqrt{2} \pi  \alpha _s}. \\
\end{split}
\end{equation}
It is also known that the classical saddle point $\frac{\diff \mathcal{I}}{\diff t'}=0$  determines the equilibrium solution as
\begin{equation}\label{eq:S_action_equilibrium_Delta_deriva}
\begin{split}
\frac{\diff (\mathcal{I}/\beta)}{\diff t'} = -\frac{\sqrt{2} J t'}{ \alpha_s}
+ \frac{h_{\text{tol}}}{8 (2\pi \alpha_s^2)^{1/4} } \frac{1}{\sqrt{t'}}
+ \frac{ \Delta  J \left(t' \epsilon \, \mathrm{csch}\left(\frac{t'
		\epsilon }{\sqrt{2} \alpha _s}\right)+\sqrt{2} \alpha _s \log \left(\tanh
	\left(\frac{t' \epsilon }{2 \sqrt{2} \alpha _s}\right)\right)\right)}{4 \pi 
	\alpha _s^2} = 0. \\
\end{split}
\end{equation}
After the asymptotic expansion and assuming $t' \ll 1$, we can obtain equilibrium solution as
\begin{equation}\label{eq:S_action_equilibrium_h_effect}
t'=
\begin{cases}
\frac{1}{4} \left( \frac{\alpha_s h_{\text{tol}}}{J} \right)^{2/3} \left( \frac{1}{8 \pi \alpha_s^2} \right)^{1/6}  & \Delta = 0,\\
\frac{\Delta}{4 \pi} \log \left( \frac{- e \epsilon \Delta}{8\sqrt{2} \pi \alpha_s} \right)
+ \frac{\Delta}{4 \pi} \log \left( -\log \left( \frac{- e \epsilon \Delta}{8\sqrt{2} \pi \alpha_s} \right)\right)
+ \cdots & h_{\text{tol}} = 0.\\
\end{cases}       
\end{equation}
The $\Delta=0$ case reproduces the result in Ref. \cite{MQ}. It is known that, in the gravity interpretation, the energy gap in the TW phase is $E_{\mathrm{gap}} = \frac{J}{\sqrt{2} \alpha_s} t' \mathcal{D}$. Then after using \eqref{eq:S_action_equilibrium_h_effect}, we reveal different contributions of $h_{\text{tol}}$ and $\Delta$ to the  $E_{\mathrm{gap}}$ term.

Second, with the value of $t'$, the magnetization $M^x$ can also be calculated by the Green's function:
\begin{equation}
\begin{split}
M^x 
&= \left| G_{\uparrow\downarrow}(t_l({\tau}),t_r({\tau}))  \right|  \\
&= \left( \frac{1}{ 16 \pi J^2 } \right)^{1/4}  \left[\frac{t_{l}^{\prime}({\tau}) t_{r}^{\prime}({\tau})}{\cosh^{2}  \frac{ \left( t_{l}({\tau})-t_{r}({\tau}) \right) }{2}}\right]^{1/4} \\
%&= \left( \frac{1}{ 16 \pi J^2 } \right)^{1/4}  \left( \frac{\diff t}{\diff \tau} \right)^{1/2} \\
&= \frac{1}{2 (2\pi)^{1/4}} \frac{\sqrt{t'}}{\sqrt{\alpha_s}}. \\
\end{split}
\end{equation}
plugging Eq.\eqref{eq:S_action_equilibrium_h_effect} into the above relation between $M^x$ and $t'$, the effects of $h_{\text{tol}}$ and $\Delta$ are shown clearly. 

\subsection{Quench Dynamics with Molecular Field Correction}

In the subsection A, we have studied the equilibrium properties by the imaginary time approach. In this subsection, we will study the real time evolution of this random spin model, starting from the TW phase. Therefore, we perform the analytically continuation on the action \eqref{eq:S_lowE_action_full} by substituting the coordinate time $\tilde{\tau} \to \iu \tilde{t} + 0 \ (\tau \to \iu t + 0)$ and reparametrization modes $t_{l/r} \to \iu t_{l/r}$. Here, for simplicity, we consider the case with $\Delta=0$. Subsequently, Eq.~\eqref{eq:S_lowE_action_full} becomes
\begin{equation} \label{eq:S_realT_action}
\begin{split}
\frac{\mathcal{I}}{N}&= \int d \tilde{t}
\Bigg\{-\left(\left\{\tan \frac{ t_{l}(\tilde{t})}{2}, \tilde{t}\right\}+\left\{\tan \frac{ t_{r}(\tilde{t})}{2}, \tilde{t}\right\}\right) + \tilde{h} \left[\frac{t_{l}^{\prime}(\tilde{t}) t_{r}^{\prime}(\tilde{t})}{\cos ^{2}  \frac{\left( t_{l}(\tilde{t})-t_{r}(\tilde{t}) \right) }{2}}\right]^{\mathcal{D}}
\Bigg\}, \\
\end{split}
\end{equation}
which is the same as the one in the Maldacena-Qi model. The action has $\operatorname{SL}(2)$ gauge symmetry, which means we should require the Noether charge vanishes $Q_0 = 2N e^{-\varphi} [-e^{2\varphi} - \varphi'' + \tilde{h} \mathcal{D} e^{2 \mathcal{D} \varphi}]=0$ with $t_l'(\tilde{t})=t_r'(\tilde{t})=e^{\varphi(\tilde{t})}$\cite{MQ}. This determines the corresponding quench dynamics, during which the magnetization at coordinate time $t$ can be expressed as
\begin{equation}
\begin{split}
M^x({t}) 
&=  \left| G_{\uparrow\downarrow}(t_l(t),t_r(t)) \right|  \\
&= \left( \frac{1}{ 16 \pi J^2 } \right)^{1/4}  \left[\frac{t_{l}^{\prime}(t) t_{r}^{\prime}(t)}{\cos^{2}  \frac{ \left( t_{l}(t)-t_{r}(t) \right) }{2}}\right]^{1/4} \\
&= \frac{1}{2 (2\pi)^{1/4} \sqrt{\alpha_s}}   e^{\frac{1}{2} \varphi({t})}. \\
\end{split}
\end{equation}
Considering the non-zero $\bar{J}$ effect in the dynamical process, we can substitute $h_{\text{tol}}$ as $h_{\text{tol}}(t) = - \bar{J} M^x({t}) $. Consequently, the equation of motion becomes 
\begin{equation}\label{eq:EOM_lowE}
-e^{2\varphi} - \varphi'' -  \frac{\bar{J}}{32 \sqrt{\pi} J}  e^{\varphi} = 0,
\end{equation}
with the proper initial conditions for this ordinary differential equation given by $\varphi(0) = 2 \log \left( 2 (2\pi)^{1/4} \sqrt{\alpha_s} M^{x}(0) \right) $ and $\varphi'(0) = 0 $. By numerically solving this differential equation, we find that for $\bar{J} < 0$ and $h>0$, and in a parameter range of $\bar{J}/J$,  $M^x$ first decreases and then oscillates around certain non-zero value as time increases.

\section{Phase transition}

Here we introduce more details of two methods that we used to determine the phase diagram. The first one is the imaginary-time approach with which we can directly evaluate the thermodynamical quantities, such as the free energy $f = -\frac{1}{\beta N} \log Z = \frac{1}{\beta N} \mathcal{I}$ (see Eq.~\eqref{eq:action_origin}) and the magnetization $M^x$. The second one is the real-time approach, from which we can obtain the spectral function $\rho_{ss'}(\omega)$($s,s'=\uparrow, \downarrow$) and the spectral ratio we introduced as $r \equiv \rho_{\uparrow\uparrow}(0) /\underset{\omega}{\operatorname{max}} \  \rho_{\uparrow\uparrow}(\omega)$.

With the imaginary-time approach, we start the self-consistent calculation from the high-temperature region, initialized with a proper high-temperature initial guess of $G(\tau)$. When the self-consistent solution is reached at a certain temperature, we decrease the temperature and continue the self-consistent calculation for a slightly lower temperature, using the current self-consistent $G(\tau)$ as an initial guess. In this way, we can obtain a series of $M^x$ and $f$ evolving from the high-temperature side, denoted by $M^x_\text{H}$ and $f_\text{H}$. The calculation can be carried out similarly by evolving from the low-temperature side, which can also obtain $M^x$ and $f$ denoted by $M^x_\text{L}$ and $f_\text{L}$. 

\begin{figure*}[tb]
	\centering
	\includegraphics[width=1.0\linewidth]{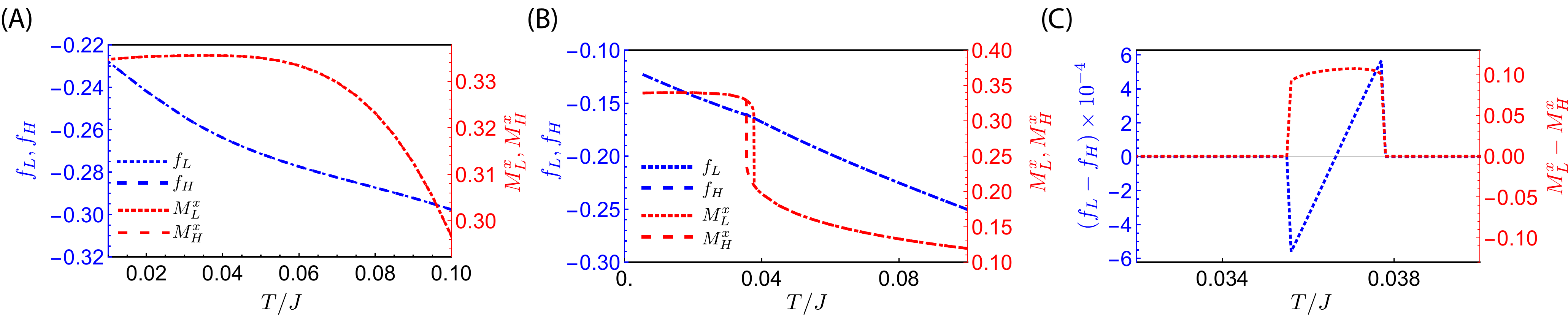}
	\caption{\textbf{Phase Transition Probed by the Imaginary Time Approach.} (A,B) $f_{L/H}$ and $M^x_{L/H}$ denote the free energy and magnetization obtained from the initialization starting from the low-temperature side and the high-temperature side respectively. Here $h_{\text{tol}}=0.2 J$, and $\Delta=-0.73$ for (A) and $\Delta=0$ for (B). (C) The difference between $f_{L/H}$ and between $M^x_{L/H}$ for (B). }
	\label{fig:S_phasediagramalphat}
\end{figure*}

After obtaining the thermodynamics data by self-consistent calculation initiated from high- and low-temperature sides, we can determine the two phases and their boundary, as well as the order of phase transitions. For the imaginary time approach, we find both the crossover behavior (Fig.~\ref{fig:S_phasediagramalphat}(A)) and the first-order transition behavior (Fig.~\ref{fig:S_phasediagramalphat}(B, C)). At relatively large $|\Delta|$ region $|\Delta| \gtrsim -0.5$, both $M^x$ and $f$ change continuously as lowering the temperature, and solutions initialized from high-temperature or low-temperature are identical. At relatively small $|\Delta|$ region, say $ |\Delta| \lesssim 0.5$, thermodynamical properties behave as typical first-order transitions. Both $M^x$ and $f$ show hysteresis effect for two different initializations from high-temperature and low-temperature sides, and a discontinuous jump exists at a certain temperature. Consequently, we can exactly determine the location of the first order transition temperature $T^*$ by the point $f_\text{H}(T^*)-f_\text{L}(T^*)=0$.

With the real-time approach, we can obtain the spectral function, and by definition, the spectral ratio $r \in [0, 1]$. $r = 0$ means a gap at $\omega=0$, describing the spectral function of the TW phase as discussed in the main text. $r=1$ means that the spectral function is peaked at $\omega=0$, describing the spectral function of the BH phase. Here we can also find different transition types. For the first order transition region, $r$ jumps discontinuously between $0$ and $1$. And in the crossover region, the gap closes gradually and $r$ changes continuously.

\section{Quench Dynamics}

\begin{figure*}[tb]
	\centering
	\includegraphics[width=0.85\linewidth]{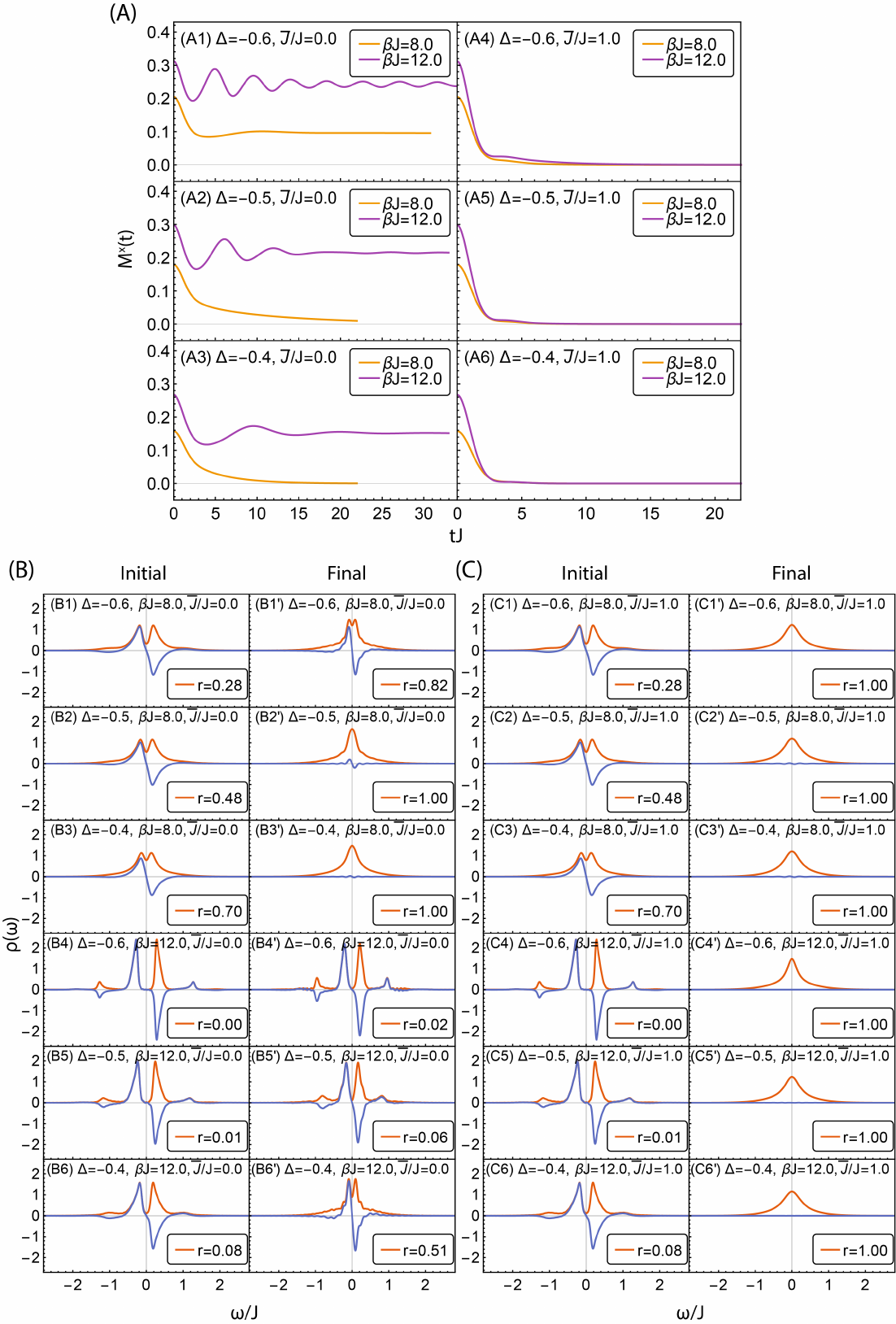}
	\caption{\textbf{More Examples of the Quench Dynamics.} (A) The time dependence of the magnetization for various parameters. (B) Spectral functions for both the initial and the final states with $\bar{J}/J=0$. (C)Spectral functions for both the initial and the final states with $\bar{J}/J=1$. }
	\label{fig:S_quenchdynamicsdetails}
\end{figure*}

In this section, we provide more examples to support the two main conclusions on the quench dynamics proposed in the main text.  Fig.~\ref{fig:S_quenchdynamicsdetails} shows the results of the quench dynamics, including the time dependence of the magnetization and the spectral functions in the initial and final states before and after the quench dynamics. 
In addition, we summarize the examples in the Table.~\ref{tab:S_quench_dynamics_data}. Clearly, in all these examples, we can see that 
\begin{itemize}
	\item When initially $r<1$ and finally $r=1$, we observe a stretched exponential decay with exponent $\eta<1$.  
	\item When initially $r<1$ and finally $r$ is also $<1$, we observe that the magnetization saturates at finite value at long time and oscillatory behavior at long time.  
\end{itemize}

\begin{table}[htbp]
	\centering
	\caption{\textbf{Summary of the Examples on Quench Dynamics.} Results with various parameters $\bar{J}/J,\ \Delta, \ \beta J$ are summarized in the table. For each given parameter, the results are described with a (decay exponent $\eta$/O,  $\text{initial } \beta J \to \text{final } \beta J$, $\text{initial } r \to \text{final } r$). Where the symbol O stands for the oscillatory behavior around a non-zero saturation value.  Here we recall that spectral ratio $r \in [0, 1]$ and $r \to 0$ means the TW phase and $r\to1$ means the BH phase.   }
	\begin{tabular}{|c|c|c|c|c|}
		\hline
		         $ \bar{J}/J $          &                         \multicolumn{2}{c|}{0}                          &                          \multicolumn{2}{c|}{1}                           \\ \hline
		$\Delta$ $\backslash$ $\beta J$ &                  8                  &                12                 &                 8                  &                  12                  \\ \hline
		             -0.6               &  O, $8.0 \to 15.3, 0.28 \to 0.82$   & O, $12.0 \to 16.6, 0.00 \to 0.02$ & 0.79, $8.0 \to 5.5, 0.28 \to 1.00$ & 0.49, $12.0 \to 8.4, 0.00 \to 1.00$  \\
		    \cmidrule{2-5}    -0.5      & 0.59, $8.0 \to 12.4, 0.48 \to 1.00$ & O, $12.0 \to 18.6, 0.01 \to 0.06$ & 0.89, $8.0 \to 5.0, 0.48 \to 1.00$ & 0.75, $12.0 \to 5.3, 0.01 \to 1.00$  \\
		    \cmidrule{2-5}    -0.4      & 0.86, $8.0 \to 8.2, 0.70 \to 1.00$  & O, $12.0 \to 22.5, 0.01 \to 0.51$ & 0.95, $8.0 \to 4.8, 0.70 \to 1.00$ & 0.91, $12.0 \to 4.3,  0.08 \to 1.00$ \\ \hline
	\end{tabular}%
	\label{tab:S_quench_dynamics_data}%
\end{table}%

\end{document}